\PassOptionsToPackage{dvipsnames}{xcolor}
\documentclass[conference]{IEEEtran}
\IEEEoverridecommandlockouts
\usepackage{cite}
\usepackage{amsmath,amssymb,amsfonts}
\usepackage{algorithmic}
\usepackage{graphicx}
\usepackage{textcomp}
\usepackage{xcolor}
\usepackage{balance}
\usepackage[all]{nowidow}

\usepackage[hidelinks]{hyperref}
\usepackage[capitalise]{cleveref}

\newcommand{\UC}{Umbilical Choir}

\begin{document}

\title{Umbilical Choir: Automated Live Testing for Edge-To-Cloud FaaS Applications\\
    \thanks{Funded by the Bundesministerium f{\"u}r Bildung und Forschung (BMBF, German Federal Ministry of Education and Research) -- 16KISK183.}

}

\author{\IEEEauthorblockN{Mohammadreza Malekabbasi, Tobias Pfandzelter, David Bermbach}
    \IEEEauthorblockA{\textit{Technische Universit\"at Berlin \& Einstein Center Digital Future}\\
    \textit{Scalable Software Systems Research Group} \\
        \{mm,tp,db\}@3s.tu-berlin.de}
}

\maketitle

\begin{abstract}
Application users react negatively to performance regressions or availability issues across software releases.
To address this, modern cloud-based applications with their multiple daily releases rely on live testing techniques such as A/B testing or canary releases.
In edge-to-cloud applications, however, which have similar problems, developers currently still have to hard-code custom live testing tooling as there is no general framework for edge-to-cloud live testing.

With \emph{\UC{}}, we partially close this gap for serverless edge-to-cloud applications.
\UC{} is compatible with all Function-as-a-Service platforms and (extensively) supports various live testing techniques, including canary releases with various geo-aware strategies, A/B testing, and gradual roll-outs.
We evaluate \UC{} through a complex release scenario showcasing various live testing techniques in a mixed edge-cloud deployments and discuss different geo-aware strategies.

\end{abstract}

\begin{IEEEkeywords}
    Continuous Deployment, Function-as-a-Service, Edge-to-Cloud, Live Testing, A/B Testing, Canary Releases
\end{IEEEkeywords}

\section{Introduction}
\label{sec:introduction}
Users of modern cloud-based application react very sensitive to performance regressions and, more generally, Quality of Service (QoS) issues~\cite{arapakis2014impact,paper_bermbach2020_webapibenchmarking2,brutlag2008user}.
As such applications typically release new versions several times a day using Continuous Integration/Continuous Deployment (CI/CD) pipelines~\cite{schermann2016bifrost}, there is tremendous risk for incurring performance regressions frequently.
While Continuous Benchmarking, e.g.,~\cite{paper_grambow2019_continuous_benchmarking,van2012kieker,paper_grambow2020_benchmarking_microservices,paper_grambow2021_optimizing_microbenchmarks,grambow2023faasterbench,rese2024duetfaas,schirmer2024elastibench}, can help to catch some performance regressions prior to release, the standard go-to approach is live testing.
In live testing, new releases are exposed to duplicate traffic in the background, serve subsets of users, or are gradually rolled out across server clusters while monitoring QoS metrics~\cite{fabijan2018online,schermann2016bifrost}.

While these techniques have become standard in cloud environments, they remain underused in the domain of edge-to-cloud applications which suffer from unique complexity challenges such as geo-distribution and heterogeneity of compute resources~\cite{paper_bermbach2017_fog_vision}.
The main reason for this is that existing cloud-based approaches such as Bifrost~\cite{schermann2016bifrost} cannot be directly used in edge-to-cloud contexts, especially since they are not geo-aware and disregard the deployment location of software components.
This means that interested developers currently have to hard-code custom live testing tooling for their applications.

To partially close this gap, we propose \UC{}, a novel live testing framework for serverless edge-to-cloud applications.
\UC{} can work alongside any Function-as-a-Service (FaaS) platform and supports various live testing techniques, additional ones can easily be added.
With \UC{}, developers can use A/B testing, canary releases with customizable geo-aware strategies, dark launches, and gradual rollouts out of the box to systematically evaluate new application releases, thus, avoiding undesired performance regressions.

In this regard, we make the following contributions:
\begin{enumerate}
	\item We describe the design of \UC{}, a generic framework for serverless edge-to-cloud live testing (\Cref{sec:uc}).
	\item We define and discuss three basic strategies for canary releases and gradual rollouts in a geo-distributed edge-to-cloud environment (\Cref{sec:strategies}).
	\item We implement \UC{} as an open source prototype\footnote{\url{https://github.com/ChaosRez/umbilical-choir-core}} in which developers can descriptively define release plans (\Cref{sec:eval:prototype}).
	\item We evaluate our approach with an example application, showcasing the supported live testing techniques as part of a complex edge-to-cloud release plan (\Cref{sec:eval:exp}).
\end{enumerate}

\section{Background and Related Work}
\label{sec:background}
In this section, we give an overview of foundations and discuss related approaches such as Bifrost~\cite{schermann2016bifrost}.

\subsection{Serverless Edge-to-Cloud Computing}
Serverless computing or FaaS is a programming model in which developers deploy individual stateless functions on a FaaS platform where the provide manages the entire function lifecycle including scaling, failure handling, etc.
While serverless computing has mainly been used in the cloud so far, the idea of serverless edge-to-cloud computing has been proposed several times, e.g.,~\cite{raith2023serverless,paper_bermbach2017_fog_vision,paper_bermbach2021_cloud_engineering}.
There are now multiple FaaS platforms natively supporting the edge-to-cloud continuum, e.g., Lean OpenWhisk~\cite{baldini2017serverless}, tinyFaaS~\cite{paper_pfandzelter2020_tinyfaas}, AuctionWhisk~\cite{paper_bermbach2021_auctionwhisk}, NanoLambda~\cite{george2020nanolambda}, Serverledge~\cite{russo2023serverledge}, or GeoFaaS~\cite{malekabbasi2024geofaas}.

\subsection{Multi-Phase Live Testing}
Multi-Phase Live Testing combines various live testing practices such as canary releases, dark launches, A/B tests, and gradual roll-outs~\cite{rahman2015synthesizing}.
These strategies are key in continuous delivery to test changes in production with minimal risk.
In canary releases, a new version is gradually rolled-out to more and more machines allowing developers to monitor its performance before a full roll-out.
In gradual roll-outs, however, a new version is rolled-out to all machines directly but only delivered to a subset of users.
This is often combined with canary releases.
In dark launches, new functionality is deployed in production serving full user traffic but the results are not delivered to end users.
This allows developers to evaluate new releases under realistic conditions while not risking customer irritation.
Finally, A/B testing deploys multiple versions of the same piece of software to distinct user groups in parallel.
The goal here is to compare the different versions regarding different KPIs such as performance but possibly also business metrics such as sale conversions.

All four are widely used for cloud applications~\cite{cito2014identifying,begel2014analyze,kim2016emerging,tang2015holistic}, supported by approaches such as Bifrost~\cite{schermann2016bifrost}.

\subsection{Testing (Serverless) Edge-to-Cloud Applications}
Serverless computing introduces distinct challenges for software testing, stemming from its dynamic behavior and the intricate interactions between functions and external services.
The event-driven architecture and inherently distributed nature of serverless applications necessitates the adoption of automated CI/CD pipelines to ensure reliability and efficiency~\cite{raith2023serverless}.
On the edge, such challenges are further aggravated~\cite{raith2023serverless,DeSilva2024The}.
To our knowledge, there is currently no consensus on best practices for testing and debugging serverless applications, particularly those spanning the edge-to-cloud continuum.
This challenge is compounded by the lack of tools and the unfamiliarity of test engineers with serverless environments, making comprehensive testing particularly difficult~\cite{DeSilva2024The, rinta2022testing}.

While research has explored the profiling and debugging of serverless function systems~\cite{shahrad2019architectural,grambow2023befaasextended}, fine-grained performance change detection in FaaS function releases~\cite{grambow2023efficiently, rese2024duetfaas}, distributed tracing for serverless applications~\cite{borges2021faaster}, and multi-cloud releases~\cite{khochare2023xfaas}, these efforts are not specifically tailored for edge-to-cloud serverless releases.
Existing approaches from the field of multi-phase live testing for microservices~\cite{schermann2016bifrost,schermann2,tarvo2015canaryadvisor} cannot be applied directly -- neither to edge-to-cloud scenarios nor to serverless applications.

Alternative approaches include creating emulated testing environments in the cloud, e.g.,~\cite{paper_hasenburg2021_mockfog2,beilharz2021continuously,paper_pfandzelter2022_celestial,behnke-hector}.
Such approaches, however, still leave the issue of rolling out the release and will usually not reach the level of confidence which can be achieved through live testing.
As such, they should be seen as complementary, running in an earlier testing stage.

\section{Design of Umbilical Choir}
\label{sec:uc}
In this section, we start by giving an overview of the high-level architecture (\cref{sec:uc:arch}) and key components (\cref{subsec:components}) of the \UC{} platform, before describing how developers can specify the release strategies for \UC{} (\cref{sec:uc:live}).

\subsection{Architecture}
\label{sec:uc:arch}
\begin{figure}[t]
    \centering
    \includegraphics[width=\linewidth]{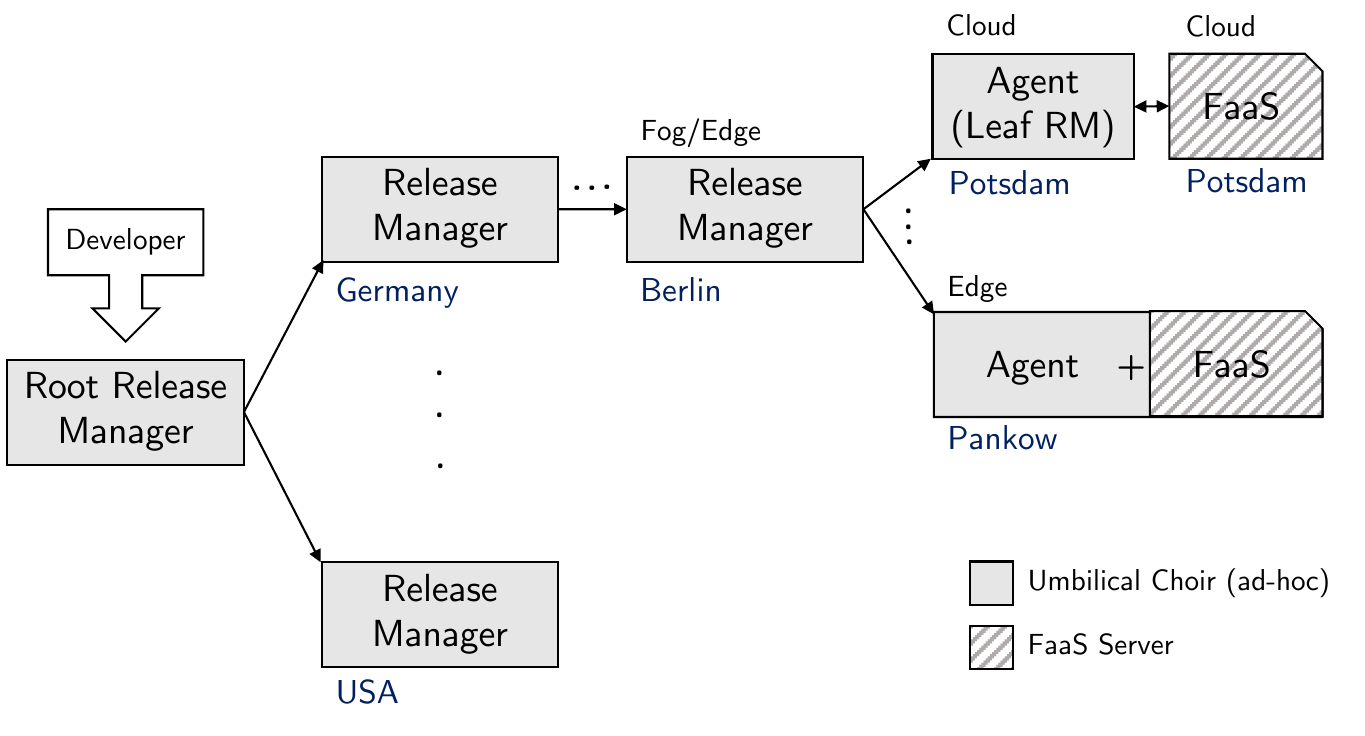}
    \caption{High-level Architecture of Umbilical Choir.}
    \label{fig:architecture}
\end{figure}

To manage the global scale and geo-distribution of application functions, \UC{} is organized in a tree-like hierarchy of nodes, with each node in the hierarchy managing a specific geographical area as shown in \cref{fig:architecture}.
Such nodes are either \textit{Release Managers} (RMs) or \textit{Agents} if they are leaf nodes.
This tree architecture was selected to provide inherent scalability, maintain operational simplicity, and enable effective delegation and aggregation of tasks, crucial for managing and monitoring geographically distributed functions.
Both node types have different functionality but provide the same interface to higher-levels, i.e., to an RM it is irrelevant whether it manages a group of Agents or of RMs.
In either case, the RM will delegate parts of a release strategy to its child nodes.
If a child node is an RM, it will take the received release strategy, devise a plan on how best to execute it, and then delegate the execution of the plan to its child nodes, each of which is responsible for a part of its parent's geographical area.
Agent child nodes in contrast will not delegate the plan further but rather handle the plan execution.
For this, they configure and deploy a \textit{proxy function} which filters client traffic, collects QoS metric data, and sends that data back to the Agent.
The collected data is then aggregated and passed up the tree towards the root RM, with each RM making decisions based on its parent's instructions and collected data.

For applications deployed in a single cloud region only, a single RM will usually suffice.
FaaS applications deployed across multiple regions or edge-to-cloud applications, will use a multi-level hierarchy of RMs.
The depth of the hierarchy depends on the number of geo-distributed FaaS locations and the capabilities of machines running the RMs.
A lower depth will be faster to run release strategies and will do it with less overhead, a higher depth level means that \UC{} can run on weak compute nodes or as a co-deployed background job on existing machines.
Since \UC{} itself is ad-hoc, it adapts to the specific needs of each application, requiring careful fine-tuning to balance performance and resource constraints.
As \cref{fig:architecture} implies, UC{} can be seamlessly integrated with already running FaaS services.

\begin{figure}[ht]
    \centering
    \includegraphics[width=\linewidth]{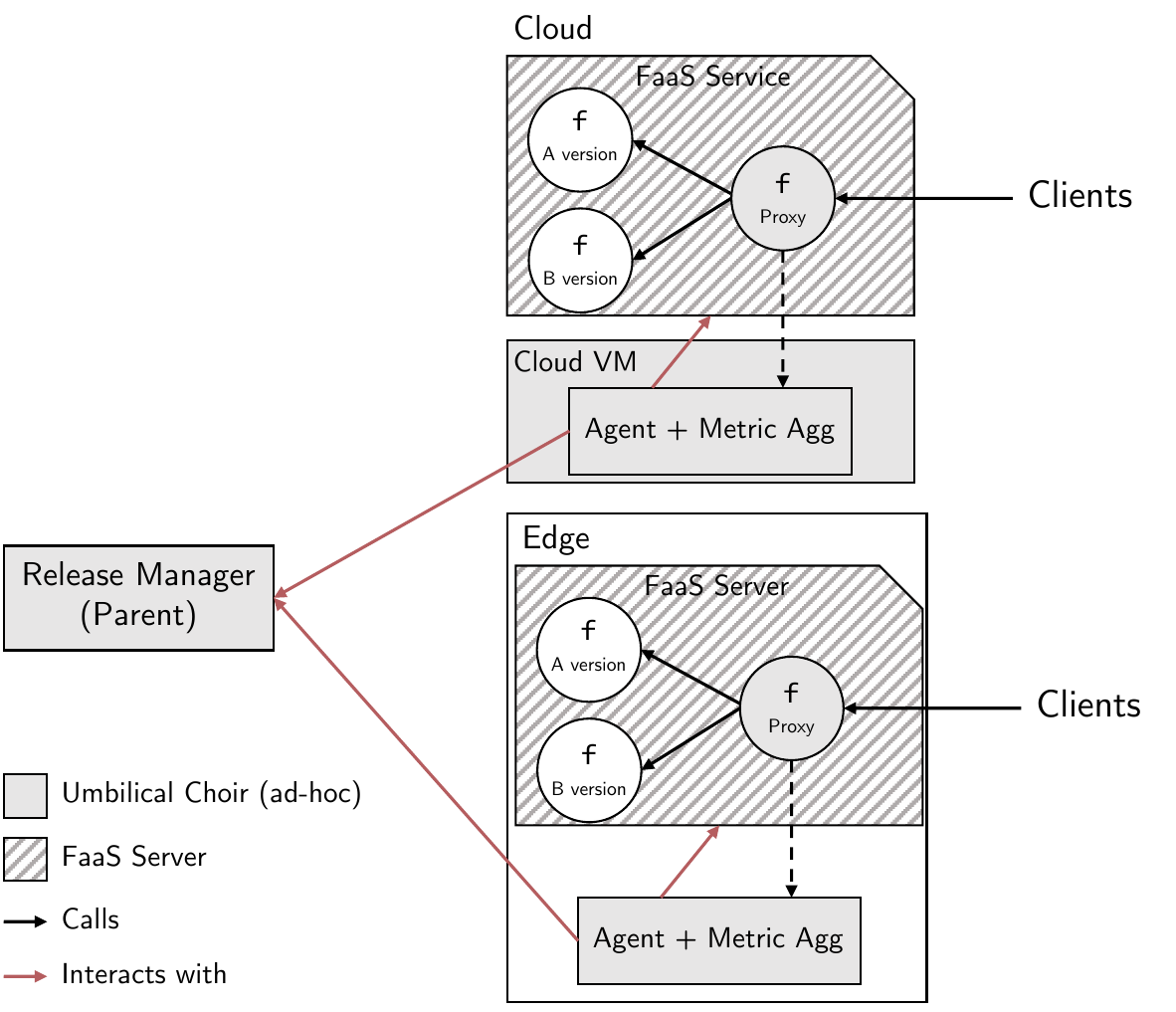}
    \caption{Agent Interaction Architecture During a Live Test}
    \label{fig:arch_agent}
\end{figure}

\subsection{Components}\label{subsec:components}
In this section, we discuss the three key components of \UC{} in more detail -- see also \cref{fig:arch_agent} for an overview how they work together.

\subsubsection{Release Manager}
RMs are organized recursively with one or more children (either RMs or Agents) and optionally a parent.
Each RM is aware of the capabilities of its child nodes, for this child RMs aggregate capabilities from their children and report them to their parent.
Furthermore, each RM is responsible for a geographic area, i.e., the area where the FaaS services on which it is deploying is running.
Often but not necessarily, it will be deployed near or on their live testing runtime targets -- on or near the edge or within the same cloud datacenter.

With this, each RM has sufficient knowledge to process release strategies.
Upon receipt of a new release strategy, either from the developer or from a parent RM, the RM plans how best to execute the release strategy with the available child nodes, their capabilities, and locations.
Based on this plan, the RM creates a new release strategy for each of its child nodes that are necessary for following the plan and forwards it to the respective nodes.

Please note that RMs are not aware of siblings or their parent's capabilities:
They have a clear area of responsibility for which they plan and execute release strategies and report results to their parent.
This is a design choice to simplify the implementation and to allow for easy scaling of the system.

\subsubsection{Agent}
While RMs delegate release strategies to their child nodes, Agents have access to (typically nearby) FaaS services which they use to execute release strategies received from their parent RM.
The Agent is responsible for deploying all function versions as well as the proxy function and then running the live test and collecting metrics.
After finishing each stage of a release strategy, the Agent checks whether the stage was successful or not according to conditions specified in the release strategy received from its parent.
In a second step, the Agent aggregates the observed metrics (including min, max, median, and mean for latency and error rates) and sends these results to its parent.

Given the heterogeneity of edge-to-cloud systems involving multiple FaaS providers, the Agent is designed to operate via a unified, provider-agnostic API.
This enables function deployment to any FaaS provider, irrespective of differences such as execution models (e.g., container-based or WASM-based).
To facilitate traffic filtering across diverse FaaS systems, the Agent deploys a proxy function to the same endpoint as the function under test, ensuring client traffic is filtered by the proxy function.

While our paper focuses on serverless edge-to-cloud live testing, it could easily be adapted to non-serverless runtimes.
For this, it would only be necessary to modify the Agent -- for instance, it could manage a Bifrost~\cite{schermann2016bifrost} instance instead of functions on a FaaS platform.

\subsubsection{Proxy Function}
To enable customized routing of client traffic to different function versions, an ad-hoc proxy function is required.
One approach integrates the proxy with multiple function versions into a single function, as demonstrated by Grambow et al.~\cite{grambow2023efficiently}.
However, this method may introduce conflicts with dependencies and functionalities, potentially leading to service disruptions or client failures.
An alternative is to manage the FaaS platform's reverse proxy for traffic control, as explored by Schermann et al.~\cite{schermann2016bifrost} in the context of microservices.
This is, however, not universally supported across FaaS providers or open-source platforms.
To address these limitations, we opted to first deploy all function under test before replacing the existing function with a proxy function which calls the functions under test as prescribed by the release strategy.
While this approach can possibly cause a short downtime during function replacement for some FaaS platforms, it offers a provider-agnostic solution that can be seamlessly applied across any FaaS platform spanning the edge-to-cloud continuum.

In the proxy function we support various routing methods to enable flexible traffic distribution between function versions as necessary for the respective release strategy:

\begin{itemize} 
    \item \emph{Client ID-based}: Traffic is routed based on the hash value of a unique client identifier (ID) supplied via the request header.
    Alternatively, the proxy can generate a universally unique identifier (UUID) via a \texttt{Set-Cookie Header}, assuming the client can store and transmit the UUID.
    This method ensures sticky sessions, assigning clients consistently to a specific version, while traffic is distributed between versions based on developer-defined percentage allocations.
    \item \emph{Header-based}: Routes traffic according to a custom \texttt{X-Function-Choice} header defined by the developer or an external service.
    This allows the developer to explicitly control the target version for each request.
    \item \emph{Random}: If no headers are defined, traffic is distributed randomly between versions based on developer-defined percentage allocations.
    Unlike the previous methods, this does not ensure sticky sessions, and a client may be routed to a different version with each request.
\end{itemize}

The request header, used to determine traffic routing, can be defined universally through, e.g., HTTP headers or gRPC metadata, ensuring compatibility with all possible FaaS platforms.

\subsection{Defining Release Strategies}
\label{sec:uc:live}
As already mentioned, a release strategy typically contains multiple \textit{Stages}.
Developers can specify for each stage
(i) the type of live testing (e.g., A/B testing or dark launch),
(ii) the regional rollout strategy (if applicable, see \cref{sec:strategies}),
(iii) a set of metric conditions which let the Agent decide whether the test was successful based on observed metrics,
(iv) a traffic filtering percentage to control the distribution of requests between the various functions under test,
and (v) the routing method as discussed above.

\section{Gradual Roll-out Strategies} \label{sec:strategies}
Canary releases and gradual roll-outs are traditionally used in non-geo-distributed systems in which all clients access a common endpoint.
In such systems, it does not matter to which subset of servers an update is rolled out first.
For geo-distributed systems, however, a novel approach is required to account for their decentralized nature.
Particularly, the geo-location of servers is likely to play an important role, e.g., when releases are tested in a subset of global markets first.
We propose the following four basic strategies for gradually rolling out updates to a group of geo-distributed servers as is typical for edge-to-cloud applications:
\begin{itemize}
    \item \textbf{Global Incremental:}
		This strategy essentially disregards geo-distribution completely.
		All FaaS platforms running a particular application, no matter where they are, start by running a small percentage, e.g., 1-5\% of the traffic against the new version.
		If this is successful they gradually increase that percentage but all FaaS platforms always run the same old/new percentage.
    \item \textbf{Local Sequential:}
		In this strategy, a single location starts running a small percentage of traffic against a new version.
		While this is successful, the location gradually increases that percentage.
		Only once 100\% has been reached, the same process starts on the next FaaS location.
    \item \textbf{Regional Incremental:}
		This strategy is a geography-aware variant of Global Incremental in which that strategy is applied to a specific region first, e.g., trying out new versions on the British market first.
		The percentage served by the new version is again identical across all FaaS locations in that geographic region.
		\item \textbf{Regional Sequential:}
		This strategy is a geography-aware variant of Local Sequential in which that strategy is applied only to a specific geographic region, e.g., gradually rolling out in a timezone first in which there is very little traffic due to night/day load patterns.
\end{itemize}
In practice, we will often find combinations of these four strategies.
Furthermore, a set of updates which are rolled out in parallel might have different per-update strategies.
Global Incremental is especially useful for quickly rolling out updates worldwide (e.g., critical security patches) at the risk of annoying a larger group of end users.
Local Sequential is the careful variant -- progress is made slowly but risk is minimized.
Both regional strategies are useful to address local peculiarities such as features which are important for one region but hardly matter for others (e.g., tsunami warnings for Japan vs. Switzerland).

These four strategies, either on their own or used in combination, allow developers to consider all possible scenarios, thus, asserting seamless and efficient roll-out choices across the edge-to-cloud continuum.
All four are supported by \UC{}.

\section{Evaluation}
\label{sec:evaluation}
In this section, we describe the results of evaluating \UC{}, our evaluation has two parts.
We first show that \UC{} can be implemented in practice through our open source proof-of-concept prototype (\cref{sec:eval:prototype}).
Second, we run a number of experiments with our prototype, both to quantify overheads as well as to demonstrate its capabilities through a use case.
For this, we describe the experiment setup (\cref{sec:eval:env}) followed by two experiments.

\subsection{Proof-of-Concept Prototype}
\label{sec:eval:prototype}
We implemented our proof-of-concept prototype in Go and made it available as open source\footnote{\url{https://github.com/ChaosRez/umbilical-choir-core}}.
The prototype currently supports the cloud platforms AWS Lambda, Google Cloud Functions, Azure Functions, and the edge FaaS platform tinyFaaS~\cite{paper_pfandzelter2020_tinyfaas}, additional ones can easily be added.
In contrast to cloud-only deployments where we can always get push access via SSH to the respective machine or FaaS platform, this may not always be an option for edge-to-cloud applications which might partially be deployed behind firewalls or in private networks.
For this reason, we decided to implement update roll-out via a pull mechanism similar to mobile phone app stores in which \UC{} child nodes periodically poll their parent for updates.
In our current prototype, RMs are registered manually with their parents, this could, however, for a production setting easily be handled by tools such as Ansible.

Release strategies are specified in the YAML format using the following fields:

\begin{itemize}
    \item \textit{Function and Version:} The file includes definitions of the functions involved in the release, along with their respective versions.
    These are used to specify traffic allocation and rollback actions.
    \item \textit{Rollback Action:} The function version to which traffic should be rolled back in case of failure or error, or as a preservative option for the developer.
    \item \textit{Stages:} Each stage of the release is detailed, including traffic routing rules for the base and new versions, as mentioned below.
    \item \textit{Stage Type:} Stages can be of different types, such as \texttt{Sequential} or \texttt{WaitForSignal}, which waits for a signal from the parent to end the stage.
    \item \textit{Function Name:} The function that is going to be tested in current stage.
    \item \textit{Traffic Allocation:} The release strategy file specifies the percentage of traffic to be routed to each function version during the stage.
    This optional field is independent from headers processed by the proxy function and serves as an alternative.
    \item \textit{Metric Conditions:} Stages may include multiple metric conditions, which are evaluated only after predefined end conditions, such as a minimum number of calls or a minimum execution duration, are satisfied.
    Metric conditions involve comparing a metric (e.g., error rate, response time) against a specified threshold.
    If the condition is met, the stage is considered successful; otherwise, it is deemed a failure.
    \item \textit{Stage Actions:} Each stage defines \texttt{onFailure} and \texttt{onSuccess} actions, allowing \UC{} Agent to proceed to the next stage or to trigger rollback/roll-out actions as needed.
\end{itemize}

At runtime, \UC{} nodes periodically call their parent's \texttt{/poll} endpoint.
When child nodes are registered to participate in a Release Strategy, that request results in their respective status codes being set to \texttt{Todo} and all stage statuses being set to \texttt{Pending}.
This triggers the child node to call its parent's \texttt{/release} endpoint to get the instructions which then changes the Release Strategy status to \texttt{Doing} and the first stage's status to \texttt{InProgress}.
In the following, the child node executes each stage and sends the results to its parent's \texttt{/result} endpoint and updates the stage's status to either \texttt{Completed}, \texttt{Failure}, or \texttt{Error}.
Once a stage is marked as \texttt{Completed}, the next stage is set to \texttt{InProgress}.
This is continued until all stages have been successfully completed or until some failure situation.
Aside from time- and count-based metrics, there is also an option (\texttt{WaitForSignal}) in which the client will poll its parents \texttt{/end\_stage} endpoint for a termination signal.
See also Tables \ref{tab:release_status} and \ref{tab:stage_status} for a complete overview of release strategy and stage status codes.

On the Agent side, the proxy functions report the results of every call including execution duration and error status to its Agent node.

\begin{table}[t]
    \centering
    \caption{State of a Release Strategy for a child}
    \begin{tabular}{|l|l|}
    \hline
    \textbf{ReleaseStatus} & \textbf{Description} \\ \hline
    \texttt{No} & The child should not get this release \\ \hline
    \texttt{Todo} & Marked to get the release \\ \hline
    \texttt{Doing} & The child is notified of the release \\ \hline
    \texttt{Done} & The child has completed all stages \\ \hline
    \texttt{Failed} & The release has failed \\ \hline
    \end{tabular}
    \label{tab:release_status}
\end{table}
    
\begin{table}[t]
    \centering
    \caption{State of a Stage of a Release Strategy for a child}
    \begin{tabular}{|l|l|} 
    \hline
    \textbf{StageStatus} & \textbf{Description} \\ \hline
    \texttt{Pending} & Stage is waiting to get started \\ \hline
    \texttt{InProgress} & Stage is in progress \\ \hline
    \texttt{WaitForSignal} & Waiting for parent to end stage \\ \hline
    \texttt{ShouldEnd} & Child should terminate the current stage \\ \hline
    \texttt{Completed} & Stage completed successfully \\ \hline
    \texttt{Failure} & Stage failed the specified criteria \\ \hline
    \texttt{Error} & Stage encountered an error \\ \hline
    \end{tabular}
    \label{tab:stage_status}
\end{table}

\subsection{Experiments}
\label{sec:eval:exp}
To evaluate the \UC{}, we ran two experiments with our prototype.
In the first experiment, we measure the latency overhead of using the proxy function which will be visible to end users calling that function;
in the second experiment, we run a realistic use case example with \UC{} to demonstrate its capabilities.
In the following, we first describe the experiment setup shared by both experiments before describing both experiments in detail.

\subsubsection{Experiment Setup}
\label{sec:eval:env}
Where feasible (i.e., when direct machine access is possible), RMs are deployed on the same physical or virtual machine as the corresponding Function-as-a-Service (FaaS) function.
In scenarios where co-location on the same node is not practical (i.e., FaaS services), Agent RMs are deployed on separate virtual machines within the same region as the respective FaaS function.

In our experiments we use two Agent nodes, one deployed on a Raspberry Pi 4 with 4 GB of RAM that also runs a tinyFaaS~\cite{paper_pfandzelter2020_tinyfaas} instance, and one running on a Google Cloud Platform (GCP) VM with 2 vCPUs and 4 GB of RAM (``e2-medium'').
Both are located in the Berlin area.
Functions are deployed on the above mentioned tinyFaaS node and Google Cloud Functions (region ``europe-west10-b'', i.e., also in the Berlin area).
We run then parent RM and emulated clients on a MacBook M2 on the same local network as the Raspberry Pi.
For the clients, we use Locust\footnote{\url{https://locust.io}} with several client IDs.
The ping delay between local network and the cloud VM is around 20ms.
We configure \UC{} to have Agents (i) poll the parent RM for new release strategies and (ii) check for completed end conditions both once per second.

\subsubsection{Experiment 1: Latency Overhead of Proxy Functions}
As discussed above, we use a co-deployed proxy function for request routing between old and new versions.
An alternative would be to wrap both function versions inside a single function artifact as done in, e.g.,~\cite{rese2024duetfaas}, which we decided not to do so as not to affect results due to the presence of several code versions within the same function instance.
Ideally, though, this would be supported by the FaaS platform natively.
As long as that is not the case, clients will suffer from additional latency from an extra hop as their function call is routed through the proxy function first.
In this experiment, we quantify that latency overhead.

For this, we deploy only one function version plus the proxy function.
We call both the function directly as a baseline and indirectly (via the proxy function using the request's header for deterministic routing) and measure call latency.

\begin{figure}
    \centering
    \includegraphics[width=\linewidth]{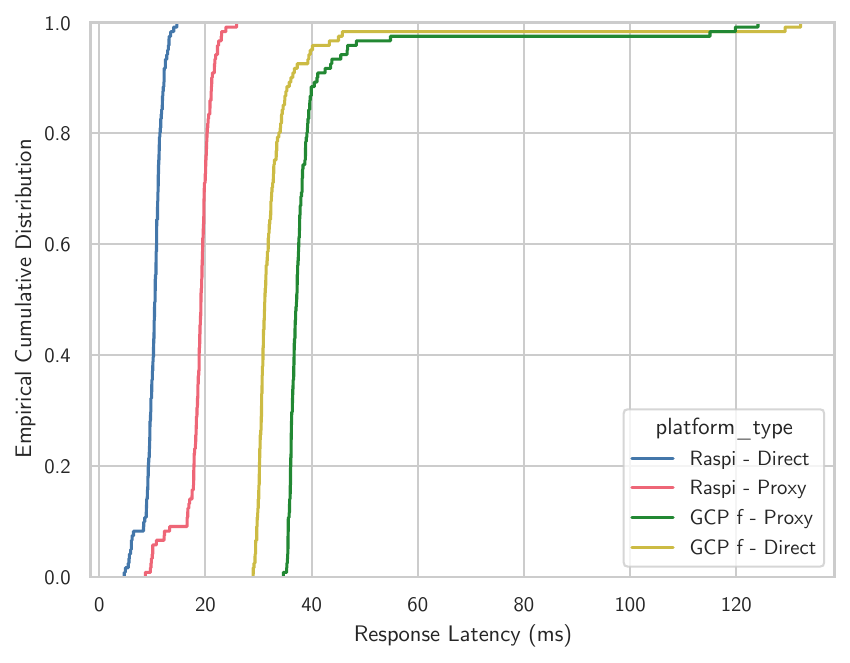}
    \caption{Response times for direct function calls and calls mediated by a proxy function, evaluated on Google Cloud Functions and Raspberry/tinyFaaS}
    \label{fig:proxy_overhead}
\end{figure}

Through pre-experiments, we asserted that load levels do not affect latency measurements.
The results which we report in \cref{fig:proxy_overhead} are from an experiment run (twice repeated) in which we send two requests per second for one minute.
As can be seen in the figure, the proxy overhead is low when Agent and both functions are deployed on the same machine whereas it is significantly higher in the Google cloud deployment.
The reason for this is that FaaS providers often have architectures with multi-hop routing for function-to-function calls~\cite{baldini2017serverless, paper_schirmer2022_fusionize, qi2024high}.
Still, we consider this overhead acceptable as long as there is no alternative such as native live testing support from the cloud provider.
The only other option we considered is co-deploying the proxy function code as part of the function (if it even is a function) calling the target function in the actual application.
This, however, introduces significant complexity, e.g., from multi-language applications, and may not always be possible.

\begin{figure*}[t]
    \centering
    \includegraphics[width=\textwidth]{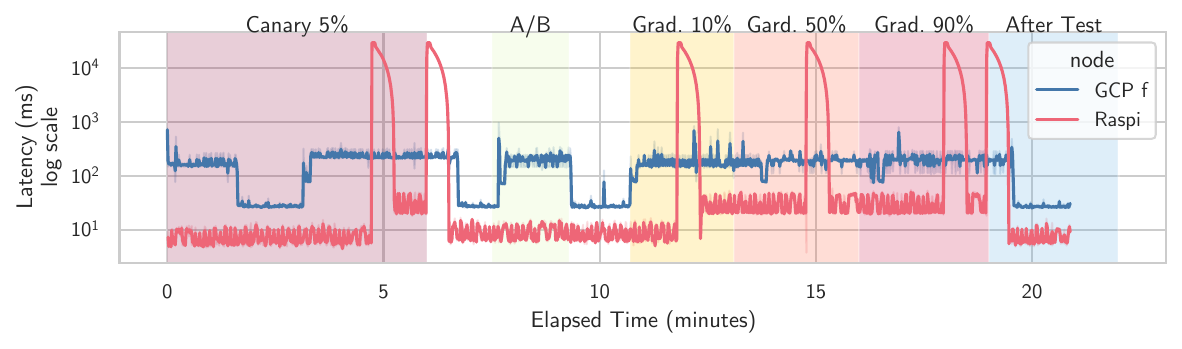}
    \caption{A realistic release strategy scenario demonstrating canary tests, A/B tests, and gradual roll-out stages while two parallel clients keep calling the function with a constant rate. The figure shows the client-observable latency over time, with one client calling the local edge node and one calling the cloud function.}
    \label{fig:realistic_scenario}
\end{figure*}

\subsubsection{Experiment 2: A Realistic Scenario}
This scenario demonstrates a realistic multi-stage release strategy for a new function version, including canary test, A/B test, and gradual roll-out.
We run one client each per FaaS location which calls the respective function twice per second.
Each stage runs for a minimum of 100 calls and 10 seconds before continuing to another stage.

In the first stage, the new version is deployed on Google Cloud Functions (GCP in \cref{fig:realistic_scenario}) for initial stability verification, with the proxy exposing 5\% of client requests randomly to the canary (new) version.
Once stable, the canary deployment is extended to the edge node running tinyFaaS.
The whole process is considered as a stage and is monitored for server errors and execution times, with the deployment being rolled back with a failure status if any of metrics should exceed a specified threshold.

The second stage implements an A/B test on the cloud side, splitting traffic 50-50 based on client IDs, i.e., sticky sessions.
We use execution duration for the success metric, since the new version is expected to be slightly more efficient (no externally measured metrics).

Following the successful A/B test, we follow the Global Incremental strategy and gradually direct more traffic to the new version on both FaaS platforms in parallel.
Initially, new version is exposed to 10\% of the traffic, and the percentage is increased to 50\% and 90\% in the next stages.
Here, we used only three stages for the gradual rollout for sake of simplicity.

In the end, all tests have been passed and the proxy function is replaced with the new function version.

As shown in \cref{fig:realistic_scenario}, \UC{} is able to handle complex release strategies on multiple nodes.
Higher numbers indicate requests served through the proxy function while lower numbers correspond to direct function calls.
Please note that the actual numbers do not matter as this experiment aims to demonstrate that \UC{} can automatically execute complex multi-stage release strategies.
Noteworthy, though, are the latency peaks on tinyFaaS which result from the way the two FaaS platforms handle redeployments of the same function.
On tinyFaaS the function is briefly unavailable whereas Google Cloud Functions keeps serving the old function.
Overall, we conclude that \UC{} achieves its desired goals.

\section{Discussion}
While \UC{} is a good first step towards integrated live testing solutions for edge-to-cloud FaaS applications, there are a number of issues which should be addressed in future work.

First, our focus was on FaaS edge-to-cloud applications.
There are, however, also non-FaaS applications and mixed FaaS/non-FaaS applications.
Neither of those two can currently be supported.

Second, we assumed no native live testing support in FaaS platforms with the resulting overheads of our proxy function.
Ideally, this would be supported by FaaS platforms, although, there may be FaaS platforms which already support such live testing strategies (possibly also via mechanisms not originally designed for live testing).
In this in-between state, our approach could be improved by leverage native live testing support where available and using proxy functions where not.
We have also focused on isolated functions rather than function workflows.
When using workflow engines such as~\cite{carl2024geoff}, it would be possible to integrate the proxy code inside the workflow engine.

Third, particularly near the edge, clients may be mobile, thus, interacting with different edge nodes.
In such a scenario, it is hard to correctly implement version consistency for clients -- this can only work if there is a deterministic way to assign and track client IDs.
In practice, this can, however, mean that one node serves only either the old or the new version to its clients, thus, blocking progress on the release strategy.

Fourth, using \UC{} causes additional costs and edge nodes may not always have enough resources to run two functions (plus the proxy) in parallel.
Developers need to carefully study whether this is an issue for their use case and whether the costs do not outweigh the benefits of testing new versions in the cloud only.

Despite this, we still see \UC{} as an overall good first step in the direction of edge-to-cloud live testing.

\section{Conclusion}
\label{sec:conclusion}
Modern cloud-based applications heavily rely on live testing techniques to ensure that frequent application releases do not lead to QoS regressions.
For edge-to-cloud applications, however, they remain underused due to their unique challenges such as geo-distribution, heterogeneous resource capabilities, and the lack of general-purpose live testing frameworks tailored for these settings.

Focusing on FaaS application, this paper proposed \UC{}, a novel live testing framework specifically designed for serverless edge-to-cloud applications.
\UC{} is compatible with all FaaS platforms and provides developers with built-in support for key live testing techniques, including A/B testing, dark launches, canary releases with customizable geo-aware strategies, and gradual rollouts.
Additional techniques can easily be added.
With \UC{}, developers can comprehensively evaluate new edge-to-cloud FaaS application releases with state-of-the-art live testing techniques in a fully automated way.

\balance

\bibliographystyle{IEEEtran}
\bibliography{bibliography,additional_related_works}

\end{document}